\documentclass[letterpaper]{emulateapj}
\shorttitle{The Mass of Kepler-93b}
\shortauthors{Dressing et al.}
\usepackage{epsfig}
\usepackage{amsmath}
\usepackage{rotating}
\usepackage{natbib}
\usepackage{enumerate}
\usepackage{hyperref}
\usepackage{url}
\usepackage{longtable}
\usepackage{xcolor}
\usepackage{xspace}
\def\msun{{\rm\,M_\odot}}                                                       
\def\rsun{{\rm\,R_\odot}}        
\def\mearth{{\rm\,M_\oplus}}                                                
\def\rearth{{\rm\,R_\oplus}}

\def\nharpsn{86\xspace}
\def\nthirteen{38\xspace}
\def\nfourteen{49\xspace}

\def\nhires{32\xspace}
\def\massest{$4.02\pm 0.68\mearth$\xspace} 
\def\rhoest{$6.88 \pm 1.18$  g/cc\xspace} 
\def\tcircest{75~Myr\xspace} 

\begin{document}
\title{The Mass of Kepler-93b and The Composition of Terrestrial Planets\altaffilmark{*}}
\author{Courtney D. Dressing\altaffilmark{1},
		David Charbonneau\altaffilmark{1},
		Xavier Dumusque\altaffilmark{1},
		Sara Gettel\altaffilmark{1}, 
		Francesco Pepe\altaffilmark{2},
		Andrew Collier Cameron\altaffilmark{3}, 
		David W. Latham\altaffilmark{1},
		Emilio Molinari\altaffilmark{4,5},
		St\'ephane Udry\altaffilmark{2},
		Laura Affer\altaffilmark{6},
		Aldo S. Bonomo\altaffilmark{7},
		Lars A. Buchhave\altaffilmark{1,8},
		Rosario Cosentino\altaffilmark{4}, 
		Pedro Figueira\altaffilmark{9,10}, 
		Aldo F. M. Fiorenzano\altaffilmark{4}, 
		Avet Harutyunyan\altaffilmark{4}, 
		Rapha\"elle D. Haywood\altaffilmark{3}, 
		John Asher Johnson\altaffilmark{1},
		Mercedes Lopez-Morales\altaffilmark{1}, 
		Christophe Lovis\altaffilmark{2},  
		Luca Malavolta\altaffilmark{11,12},
		Michel Mayor\altaffilmark{2}, 
		Giusi Micela\altaffilmark{6}, 
		Fatemeh Motalebi\altaffilmark{2}, 
		Valerio Nascimbeni\altaffilmark{12}, 
		David F. Phillips\altaffilmark{1}, 
		Giampaolo Piotto\altaffilmark{11,12}, 
		Don Pollacco\altaffilmark{13}, 
		Didier Queloz\altaffilmark{2,14}, 
		Ken Rice\altaffilmark{15}, 
		Dimitar Sasselov\altaffilmark{1},  
		Damien S\'egransan\altaffilmark{2},
		Alessandro Sozzetti\altaffilmark{7}, 
		Andrew Szentgyorgyi\altaffilmark{1}, 
		Chris Watson\altaffilmark{16}}
		\altaffiltext{*}{Based on observations made with the Italian Telescopio Nazionale Galileo (TNG) operated on the island of La Palma by the Fundaci\'on Galileo Galilei of the INAF (Istituto Nazionale di Astrofisica) at the Spanish Observatorio del Roque de los Muchachos of the Instituto de Astrofisica de Canarias.}
\altaffiltext{1}{Harvard-Smithsonian Center for Astrophysics, 60 Garden Street, Cambridge, Massachusetts 02138, USA; \href{mailto:cdressing@cfa.harvard.edu}{cdressing@cfa.harvard.edu}}
\altaffiltext{2}{Observatoire Astronomique de l'Universit\'e de Gen\`eve, 51 ch. des Maillettes, 1290 Versoix, Switzerland}
\altaffiltext{3}{SUPA, School of Physics \& Astronomy, University of St. Andrews, North Haugh, St. Andrews Fife,
KY16 9SS, UK}
\altaffiltext{4}{INAF - Fundaci\'on Galileo Galilei, Rambla Jos\'e Ana Fernandez P\'erez 7, 38712 Bre\~na Baja, Spain}
\altaffiltext{5}{INAF - IASF Milano, via Bassini 15, 20133, Milano, Italy}
\altaffiltext{6}{INAF - Osservatorio Astronomico di Palermo, Piazza del Parlamento 1, 90124 Palermo, Italy }
\altaffiltext{7}{INAF - Osservatorio Astrofisico di Torino, via Osservatorio 20, 10025 Pino Torinese, Italy}
\altaffiltext{8}{Centre for Star and Planet Formation, Natural History Museum of Denmark, University of Copenhagen,
DK-1350 Copenhagen, Denmark}
\altaffiltext{9}{Centro de Astrof\`{i}sica, Universidade do Porto, Rua das Estrelas, 4150-762 Porto, Portugal}
\altaffiltext{10}{Instituto de Astrof\' isica e Ci\^encias do Espa\c{c}o, Universidade do Porto, CAUP, Rua das Estrelas, PT4150-762 Porto, Portugal}
\altaffiltext{11}{Dipartimento di Fisica e Astronomia ``Galileo Galilei", Universita'di Padova, Vicolo dell'Osservatorio 3,
35122 Padova, Italy}
\altaffiltext{12}{INAF - Osservatorio Astronomico di Padova, Vicolo dell'Osservatorio 5, 35122 Padova, Italy }
\altaffiltext{13}{Department of Physics, University of Warwick, Gibbet Hill Road, Coventry CV4 7AL, UK}
\altaffiltext{14}{Cavendish Laboratory, J J Thomson Avenue, Cambridge CB3 0HE, UK}
\altaffiltext{15}{SUPA, Institute for Astronomy, University of Edinburgh, Royal Observatory, Blackford Hill, Edinburgh, EH93HJ, UK}
\altaffiltext{16}{Astrophysics Research Centre, School of Mathematics and Physics, Queens University, Belfast, UK}

\vspace{0.5\baselineskip}
\date{\today}

\begin{abstract}
Kepler-93b is a $1.478 \pm 0.019\rearth$ planet with a 4.7~day period around a bright (\mbox{$V=10.2$}), astroseismically-characterized host star with a mass of $0.911\pm0.033\msun$ and a radius of \mbox{$0.919\pm0.011\rsun$}.  Based on \nharpsn radial velocity observations obtained with the HARPS-N spectrograph on the Telescopio Nazionale Galileo and \nhires archival Keck/HIRES observations, we present a precise mass estimate of \massest. The corresponding high density of \rhoest is consistent with a rocky composition of primarily iron and magnesium silicate. We compare Kepler-93b to other dense planets with well-constrained parameters and find that between $1-6\mearth$, all dense planets including the Earth and Venus are well-described by the same fixed ratio of iron to magnesium silicate. There are as of yet no examples of such planets with masses $>6\mearth$: All known planets in this mass regime have lower densities requiring significant fractions of volatiles or H/He gas. We also constrain the mass and period of the outer companion in the Kepler-93 system from the long-term radial velocity trend and archival adaptive optics images. As the sample of dense planets with well-constrained masses and radii continues to grow, we will be able to test whether the fixed compositional model found for the seven dense planets considered in this paper extends to the full population of $1-6\mearth$ planets.
\end{abstract}


\maketitle

\section{Introduction}
Small planets are abundant in the galaxy, but the compositional diversity of small planets is not well understood. Theoretical models of planet formation predict that planets intermediate in size between Earth and Neptune could be gaseous ``mini-Neptunes,'' water worlds, or rocky ``Super-Earths'' \citep{kuchner_et_al2003, leger_et_al2004, valencia_et_al2006, seager_et_al2007, fortney_et_al2007, rogers_et_al2011, lopez_et_al2012, zeng+sasselov2013}. Recent studies have explored the compositional diversity of small planets using hierarchical Bayesian modeling of the observed planet radii and measured planet masses \citep{rogers2014} or theoretical models \citep{wolfgang+lopez2014}, but a thorough investigation of planet densities is hindered by the small number of small planets with well-measured masses and radii.  There are currently only nine planets smaller than $2.7\rearth$ with masses measured to 20\% precision: 55 Cnc e \citep{gillon_et_al2012,  nelson_et_al2014}, CoRoT-7b \citep{barros_et_al2014, haywood_et_al2014}, GJ1214b \citep{charbonneau_et_al2009}, HD97658b \citep{dragomir_et_al2013}, HIP116454b \citep{vanderburg_et_al2014}, Kepler-36b \citep{carter_et_al2012}, Kepler-78b \citep{pepe_et_al2013, howard_et_al2013}, and Kepler-10b and 10c \citep{dumusque_et_al2014}. 

The host star Kepler-93 (KIC~3544595, KOI~69) is one of the brightest stars observed by \emph{Kepler} ($V = 10.2$, $Kp =9.93$), enabling very high precision photometry of 17~ppm on six-hour timescales \citep{christiansen_et_al2012}. \emph{Kepler} observed Kepler-93 throughout the baseline mission (Quarters 0--17) and conducted observations at short cadence (exposure time of 58.5~s) beginning in Quarter~2 and extending until the end of the mission. Due to the high photometric precision of the \mbox{Kepler-93} observations, the planet was detected in the first four months of \emph{Kepler} data \citep{borucki_et_al2011b}. \citet{marcy_et_al2014} acquired \nhires~Keck HIRES radial velocity observations of Kepler-93 from July 2009 - September 2012 and provided an estimate of $2.6 \pm 2.0 \mearth$ for the mass of Kepler-93b. \citet{marcy_et_al2014} also noted a large linear RV trend of $11.2 \pm 1.5$ m s$^{-1}$ yr$^{-1}$ and calculated lower limits on the mass and period of the perturbing companion of $M > 3M_{\rm Jup}$ and $P > 5$~yr. Incorporating an additional 14~spectra from the 2013 observing season, the HIRES mass estimate for Kepler-93b increased to $3.8\pm1.5\mearth$ \citep{ballard_et_al2014}. Nonetheless, the 40\% error on the mass measurement allows a wide range of planetary compositions including a rocky body, an ice world, and even a substantial primordial envelope of hydrogen and helium \citep{ballard_et_al2014}. 

In contrast, the properties of the host star \mbox{Kepler-93} are well-constrained. Using 37~months of \emph{Kepler} short cadence data, \citet{ballard_et_al2014} conducted an asteroseismic investigation to characterize Kepler-93 in exquisite detail. They estimated an average stellar density of $1.652\pm0.006$~g~cm$^{-3}$, a stellar mass of $0.911\pm0.033\msun$, and a stellar radius of $0.919\pm0.011\rsun$. Adopting priors from their asteroseismic investigation, they fit the \emph{Kepler} photometry to obtain a precise radius estimate of $1.478\pm0.019\rearth$ for Kepler-93b. 

In addition to characterizing the host star, \citet{ballard_et_al2014} present a variety of evidence that Kepler-93b is a bona fide planet rather than an astrophysical false positive. First, they report that the steep shape of the ingress and egress portions of the Kepler-93b light curve cannot be reproduced by a non-planetary companion. Second, they note that the infrared transit depth they measured with the \emph{Spitzer Space Telescope} is consistent with the planetary interpretation of Kepler-93b. Third, they place stringent limits on the presence of nearby stars based on Keck AO images \citep{marcy_et_al2014}. Fourth, they state that the stellar density derived from the transit duration \citep{seager+mallen-ornelas2003, nutzman_et_al2011} is consistent with the asteroseismic stellar density constraint, indicating that the planet likely orbits the target star rather than the companion causing the large RV trend. 

In this paper we refine the mass measurement of Kepler-93b from $2.5\sigma$ to $6\sigma$ by analyzing two seasons of HARPS-N radial velocities in addition to the publicly available HIRES data. We discuss these observations and our data reduction methods in Section~\ref{sec:obs}. In Section~\ref{sec:rv}, we develop a model to fit the observed radial velocities. Finally, we discuss the implications of the resulting planet mass and present our conclusions in Section~\ref{sec:disc}.

\section{Observations \& Data Reduction}
\label{sec:obs}
We obtained \nharpsn~spectra of Kepler-93 using the HARPS-N spectrograph on the 3.57-m Telescopio Nazionale Galileo (TNG) at the Observatorio del Roque de los Muchachos. HARPS-N is a high-precision, vacuum-stabilized, high-resolution ($R \simeq 115,000$) echelle spectrograph. The design is very similar to the design of the original HARPS instrument at the ESO 3.6-m \citep{mayor_et_al2003}. The main differences are that HARPS-N is fed by octagonal fibers rather than circular fibers to improve the scrambling of the light and features a monolithic 4096 x 4096 CCD instead of the dual CCD configuration used for the HARPS focal plane \citep{consentino_et_al2012}. 

We acquired \nthirteen and \nfourteen~HARPS-N observations of Kepler-93 during the 2013 and 2014 observing seasons, respectively. In most cases, we used an exposure time of 30~minutes and achieved a mean S/N per extracted pixel of 103 at 550nm. (Four of the spectra had an exposure time of 15~minutes and one had 27~minutes; these were all gathered in July 2013.) One of the observations collected in 2013 was contaminated by light from a mercury lamp and was therefore removed from the analysis. The final HARPS-N dataset analyzed in this paper consists of \nharpsn~spectra. In most cases (75 of \nharpsn~spectra), we observed Kepler-93 using simultaneous thorium argon (observing mode {\tt HARPN\_ech\_obs\_thosimult}). The remaining eleven observations were obtained without simultaneous thorium argon in observing mode {\tt HARPN\_ech\_obs\_objAB}.

 \begin{deluxetable*}{ccccccc}
\tablecolumns{7}
\tablecaption{HARPS-N Radial Velocity Observations of Kepler-93}
\tablehead{
\colhead{BJD$_{\rm UTC}$} &
\multicolumn{2}{c}{RV (m/s)} &
\colhead{Bisector} &
\multicolumn{2}{c}{log($R'_{HK}$) (dex)} & 
\colhead{$t_{\rm exp}$} \\[0.2em]
\cline{2-3}
\cline{5-6}\\[-0.8em]
\colhead{-2450000} &
\colhead{Value} &
\colhead{Error } &
\colhead{(m/s)} &
\colhead{Value} & 
\colhead{Error } &
\colhead{(s)} 
}
2456462.686262  & 27335.24  &    1.02 &  -28.39   &  -5.01  &  0.01  &  1800 \\
2456463.584483  & 27337.75  &    0.94  & -31.63   & -5.00   &  0.01  &  1800 \\
2456464.609617  & 27331.57  &    1.75  & -27.24   & -5.04   &  0.02  &  1800 \\
2456465.606438  & 27342.34  &    1.00  & -32.74   & -5.02   &  0.01  &  1800 \\
2456466.608850  & 27337.58  &    0.86  & -29.26   & -5.00   &  0.01  &  1800 \\
$...$ & $...$ & $...$ & $...$ & $...$ & $...$ &  $...$ 
\enddata
\tablenotetext{}{(This table is available in its entirety in a machine-readable form in the online journal. A portion is shown here for guidance regarding its form and content.)}
\label{tab:rvs}
\end{deluxetable*}

We reduced the data with the standard HARPS-N pipeline by cross-correlating the observed spectra with a numerical mask based on the spectrum of a G2V~star \citep{baranne_et_al1996, pepe_et_al2002}. We provide the resulting RVs and their $1\sigma$ errors in Table~\ref{tab:rvs} along with the observation BJDs, exposure times, bisector spans, and stellar activity levels as measured by the Ca II $\log(R'_{\rm HK}$) activity indicator \citep{noyes_et_al1984}. The BJDs in  Table~\ref{tab:rvs} are provided in UTC, but we converted the times to TDB (the units used by the \emph{Kepler} mission) using the IDL routine {\tt utc2bjd.pro}\footnote{\url{http://astroutils.astronomy.ohio-state.edu/time/pro/utc2bjd.pro}} prior to fitting the RVs. We did not find evidence for a correlation between RV and bisector span or $\log(R'_{\rm HK})$. 

\section{Analysis of the Radial Velocity Data}
\label{sec:rv}
Our full data set included RVs from four~seasons of HIRES observations (2009 July -- 2012 September) and two seasons of HARPS-N observations (2013 June -- 2014 October). We fit the combined HARPS-N and HIRES data set by incorporating a single offset $RV_{\rm off}$ between the HIRES and HARPS-N data. We used the following general model:
 \begin{equation}
\begin{split}
\mathcal{M}(t_i)& = \gamma + \rm{RV}_{\rm off} + \beta(t_i) \\
	& \quad +K \left[\cos(\theta(t_i, T_C, P, e) + \omega) + e \cos{\omega}\right] 
\end{split}
\end{equation}
where $\gamma$ is the systemic velocity of Kepler-93, \mbox{$RV_{\rm off} = RV_{\rm HARPS-N}-RV_{\rm HIRES}$} is the offset between the HIRES and {HARPS-N} RVs, $\beta(t_i)$ is a long-term RV trend due to a third component in the system, $K$ is the semi-amplitude due to Kepler-93b, and $\omega$ is the argument of periastron. The function $\theta$ is the true anomaly of Kepler-93b at time $t_i$ and depends on the period $P$, epoch of transit $T_C$, eccentricity $e$. When fitting eccentric orbits, we used IDL routine {\tt keplereq.pro} written by Brian Jackson\footnote{\url{http://www.lpl.arizona.edu/~bjackson/idl_code/keplereq.pro}} to solve Kepler's equation for the eccentric anomaly. The routine uses the method suggested by \citet{mikkola1987} as an initial guess.

We considered linear and quadratic parameterizations of the long-term trend $\beta(t_i)$ and circular and eccentric orbits for Kepler-93b. For all models, we determined an initial solution using the Levenberg-Marquardt minimization algorithm as implemented by {\tt lmfit} in IDL. We then explored the region of parameter-space near the best-fit solution using a Bayesian Markov Chain Monte Carlo analysis with a Metropolis-Hastings acceptance criterion \citep{metropolis_et_al1953}. We initialized $N$ chains, where $N$ was twice the number of free parameters in the chosen model. We selected different initial positions for each chain by perturbing each free parameter of the best-fit solution by a random number drawn from a distribution with a width of five times the step size. We tuned the step sizes such that the acceptance fractions for each parameter were 10--30\%. For the MCMC analysis, we set uniform priors for all parameters except the orbital period and epoch of transit. We allowed only non-negative values for the RV semi-amplitude $K$ and the separate stellar jitter terms $\sigma_{sj}$ for the HIRES and HARPS-N observations (see below).  

The \emph{Kepler} photometry places tight constraints on the period and epoch of transit \citep{ballard_et_al2014}. We incorporated this knowledge into our MCMC analysis by including Gaussian priors on period and transit epoch in the likelihood calculation. As shown in \citet{dumusque_et_al2014}, using the tight prior from Kepler photometry when fitting a circular model to RV observations of a planet in a circular orbit yields a result very similar to that from a combined photometric and spectroscopic fit. We also tested fitting the data while allowing the epoch of transit to float and find the epoch of transit at \mbox{BJD = 2454944.29514}.  This epoch differs from the value determined by \citet{ballard_et_al2014} by 4 minutes ($0.3\sigma$). The possible shift in the transit center is therefore insignificant. Accordingly, we adopt the photometric ephemeris determined by \citet{ballard_et_al2014}. 

In our calculations, we shifted the epoch of transit close to the start of the HARPS-N RVs to reduce error propagation. We increased the efficiency of our model fits by parameterizing eccentric models using  $\sqrt{e} \cos(\omega)$ and $\sqrt{e} \sin(\omega)$ rather than varying $e$ and $\omega$ directly \citep{ford2006, eastman_et_al2013}. As in \citet{dumusque_et_al2014}, we accounted for stellar activity by incorporating a stellar jitter term $\sigma_{sj}$ in our adopted likelihood $\mathcal{L}$: 
\begin{equation}
\mathcal{L}= \prod_{i=1}^{N}\left({\frac{1}{\sqrt{2 \pi (\sigma_i^2 + \sigma_{sj}^2)}} \exp \left[ - \frac{(RV(t_i) - \mathcal{M}(t_i))^2}{2 (\sigma_i^2+\sigma_{sj}^2)}\right] }\right)
\end{equation}
where $RV(t_i)$ is the measured RV at each time $t_i$ in the set of $N$ observations, $\mathcal{M}$ is the model, $\sigma_i$ is the instrumental noise listed in Table~\ref{tab:rvs}, and the stellar jitter noise $\sigma_{sj}$ is allowed to adopt a different constant value for the HARPS-N and HIRES data. 

We ran each chain for a minimum of $10^4$ steps and checked for convergence by computing the Gelman-Rubin potential scale reduction factor $\hat{R}$ for each parameter \citep{gelman_et_al2004}. We stopped the MCMC analysis when $\hat{R}< 1.03$ for all parameters. Next, we accounted for ``burn-in'' by identifying the point in each chain at which the likelihood first became higher than the median likelihood of the chain and removing all earlier steps. After combining all of the chains, we selected the median values of each parameter as the best-fit value and assigned symmetric errors encompassing 68\% of values closest to the adopted best-fit value. 

We then used Bayesian statistics to determine which of the models considered best describes the data. We followed the method of \citet{chib+jeliazkov2001} as described in the Appendix of \citet{haywood_et_al2014} to calculate the Bayes factor between pairs of models using the posterior distributions and acceptance probabilities from our MCMC analyses.  This method was previously used by \citet{dumusque_et_al2014} to compare RV models of the Kepler-10 system. We found that penalties incurred by the additional complexity of fitting the orbit of Kepler-93b with an eccentric model or fitting the long-term trend with a quadratic model outweighed the improvement in the likelihood. We also compared the models by holding the stellar jitter terms fixed to $\sigma_{sj, {\rm HARPS-N}}=1.56$~m~s$^{-1}$ and $\sigma_{sj, {\rm HIRES}}=2.03$~m~s$^{-1}$ 
and computing the Bayesian Information Criterion \citep[BIC, ][]{schwarz1978} and finite sample Akaike Information Criterion \citep[AIC$\rm{_C}$,][]{hurvich+tsai1989}. When considering only the HARPS-N data, we found that the model with a quadratic trend and a circular orbit for Kepler-93b was preferred over the models with a linear trend and circular orbit (\mbox{$\Delta$BIC = 6.1}, \mbox{$\Delta$AIC$\rm{_C}$ = 8.3}), linear trend and eccentric orbit  (\mbox{$\Delta$BIC = 12.4}, \mbox{$\Delta$AIC$\rm{_C}$ = 10.4}), or quadratic trend and eccentric orbit  (\mbox{$\Delta$BIC = 6.6}, \mbox{$\Delta$AIC$\rm{_C}$ = 2.5).}

\begin{figure*}[tbp] 
\begin{center}
\centering
\includegraphics[width=0.49\textwidth]{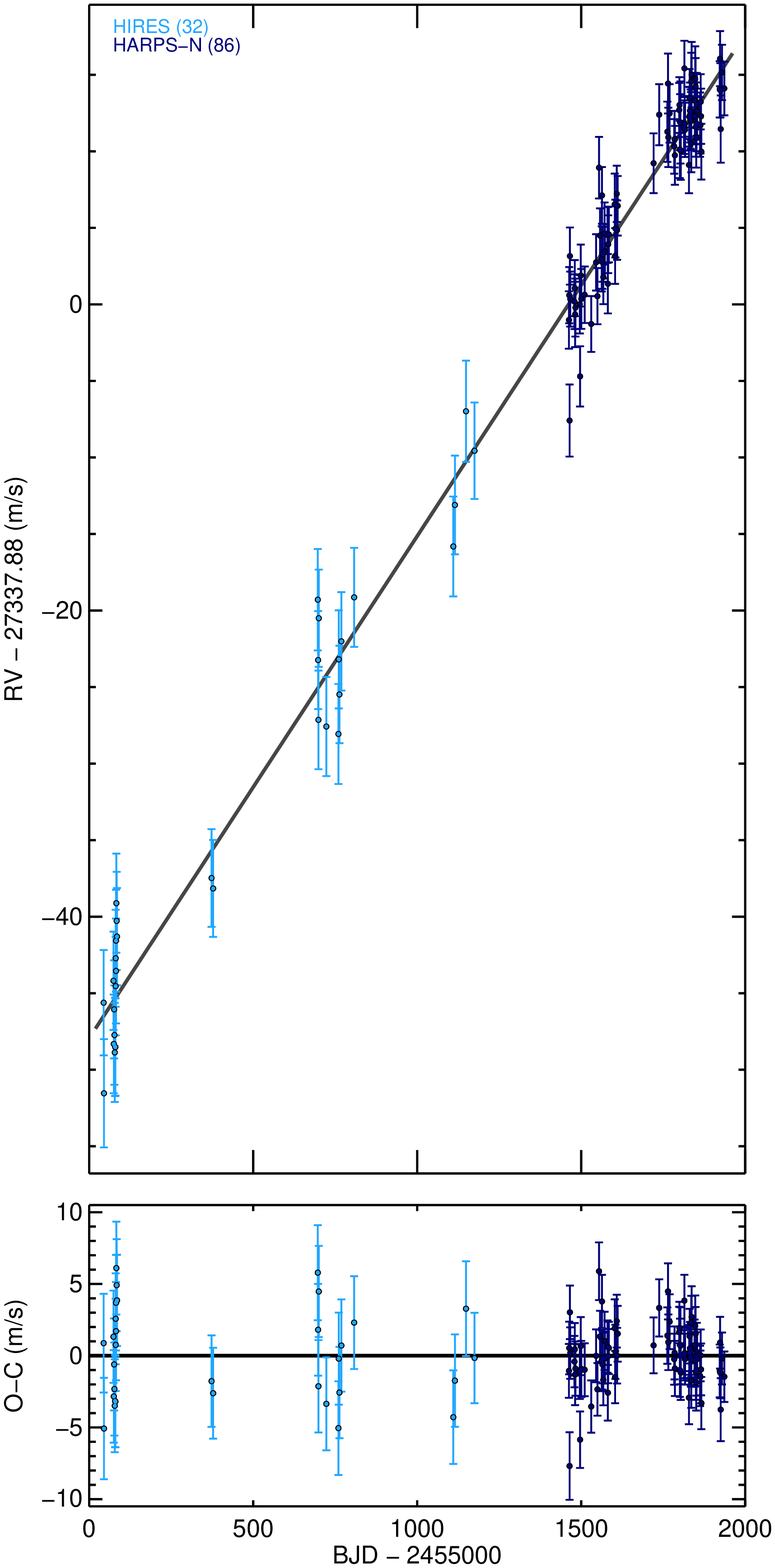}
\includegraphics[width=0.49\textwidth]{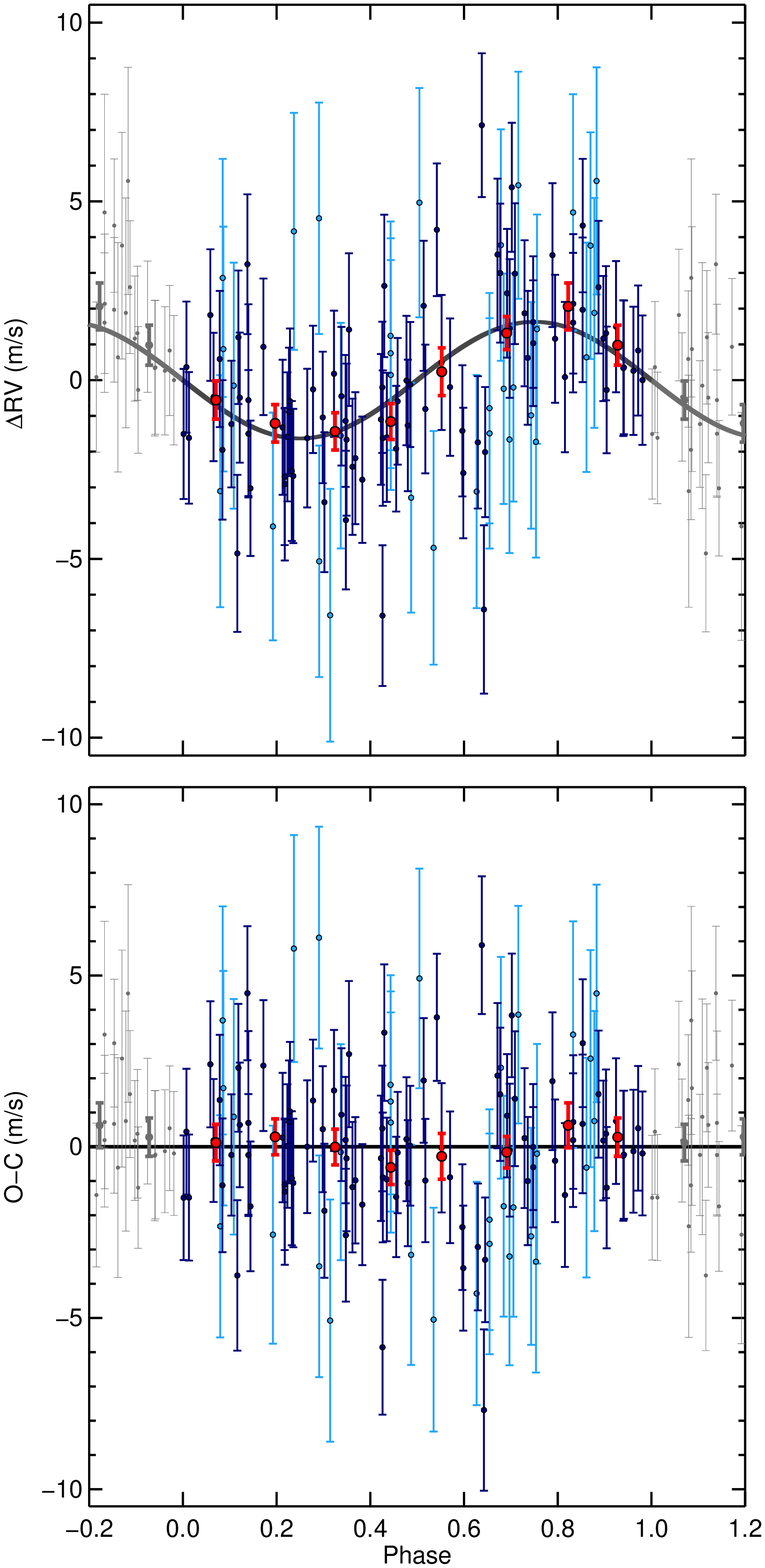}
\end{center}
\caption{Best-fit model (black) for the Kepler-93 system and measured HIRES (light blue) and HARPS-N (dark blue) RVs after correcting for the offset between HIRES and HARPS-N. The errors include contributions from both instrumental noise and stellar jitter. \emph{Top Left:} Measured RVs versus time after removing the signal of the planet \mbox{Kepler-93b}. \emph{Bottom Left: } RV residuals versus time after removing the full planet+trend fit. \emph{Top Right: } Phase-folded signal of Kepler-93b after removing the long-term trend due to Kepler-93c. The large red circles with error bars show the weighted mean and corresponding uncertainties of the measured RVs, conveniently binned to equal arbitrary intervals in phase. The points shown in gray are repeated to better reveal the behavior of the data near phase=0. \emph{Bottom Right: } RV residuals versus phase after removing the full planet+trend fit. The red circles are the binned data.}
\label{fig:rv}
\end{figure*}

\begin{figure}[tbp] 
\begin{center}
\centering
\includegraphics[width=0.5\textwidth]{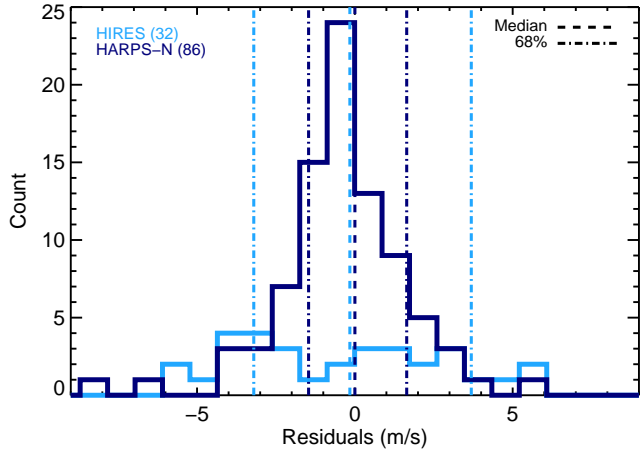}
\end{center}
\caption{Histogram of the residuals of the HIRES (light blue) and HARPS-N (dark blue) observations. The dashed lines mark the median of each distribution (-0.16 m/s for HIRES, 0.002 m/s for HARPS-N) and the dot-dash lines encompass 68\% of the measurements. The half width of the 68\% interval is 1.6~m~s$^{-1}$ for the HARPS-N data and 3.4~m~s$^{-1}$ for the HIRES data.}
\label{fig:resid}
\end{figure}

Nonetheless, when we included the HIRES data we found that the simplest model (linear trend and circular orbit) was preferred over the models with a linear trend and eccentric orbit (\mbox{$\Delta$BIC = 5.6}, \mbox{$\Delta$AIC$\rm{_C}$ = 0.7}), quadratic trend and circular orbit  (\mbox{$\Delta$BIC = 4.7}, \mbox{$\Delta$AIC$\rm{_C}$ = 2.1}), or quadratic trend and eccentric orbit  (\mbox{$\Delta$BIC = 10.5}, \mbox{$\Delta$AIC$\rm{_C}$ = 3.0}). We therefore treat the perturbation from Kepler-93c as a linear trend and model the orbit of Kepler-93b as circular. (For the eccentric fits, we found a median eccentricity of 0.15 and an upper limit of $e < 0.31$ with 95\% confidence.) We present the resulting system properties including a mass estimate for \mbox{Kepler-93b} of \massest in Table~\ref{tab:params} and display the measured RVs and the best-fit model in Figure~\ref{fig:rv}. As highlighted in Figure~\ref{fig:resid}, the HARPS-N residuals are gaussian with a distribution centered on zero and containing 68\% of the data within a half width 1.6~m~s$^{-1}$. For the HIRES residuals the region encompassing 68\% of the data has a half width of 3.4~m~s$^{-1}$.

 \begin{deluxetable}{lcc}
\tablecolumns{3}
\tabletypesize{\footnotesize}
\tablecaption{Parameters for the Kepler-93 System}
\tablehead{
\colhead{} & 
\colhead{Value and } & 
\colhead{} \\
\colhead{Parameter} &
\colhead{$1\sigma$ Errors} & 
\colhead{Ref.}
}
\multicolumn{3}{l}{\textbf{Kepler-93 (star) = KIC 3544595 = KOI 69}}\\
\hline\\[-0.7em]
Right ascension & $19^{\rm h}25^{\rm m}40.^{\rm s}39$ & 1,2 \\[0.15em]
Declination & $+38^{\rm d}40^{\rm m}20.^{\rm s}45$ & 1,2 \\[0.15em]
Kepler magnitude & 9.931 & 3 \\[0.15em]
2MASS K & 8.370 & 3 \\[0.15em]
$T_{\rm eff}$ (K) & $5669 \pm 75$ & 4 \\[0.15em]
$R_*$ (solar radii) & $0.919 \pm 0.011$ & 4 \\[0.15em]
$M_*$ (solar masses) & $0.911\pm 0.033$ & 4 \\[0.15em]
$[Fe/H]$ & $-0.18 \pm 0.10$ & 4 \\[0.15em]
$\log g$ & $4.470 \pm 0.004$ & 4 \\[0.15em]
Age (Gyr) & $6.6 \pm 0.9$ & 4 \\[0.15em]
Systemic Velocity\tablenotemark{a} (m s$^{-1}$) & $27337.89 \pm 0.51$ & 5 \\[0.15em]
HIRES Offset (m s$^{-1}$) & $27304.1 \pm 1.5$ & 5 \\[0.15em]
RV Jitter (HARPS-N) & $1.58 \pm 0.19$  & 5 \\[0.15em]
RV Jitter  (HIRES) & $2.09 \pm 0.71$  & 5 \\[0.15em]
\hline\\[-0.7em]
\multicolumn{3}{l}{\textbf{Kepler-93b (planet) = KOI 69.01}}\\
\hline\\[-0.7em]
\multicolumn{3}{l}{\emph{Transit and orbital parameters} }\\
Orbital period $P$ (days) & $ 4.72673978 \pm 9.7 \times 10^{-7}$ & 4 \\[0.15em]
Transit epoch $T_C$ (BJD) & $2454944.29227 \pm 0.00013$ & 4 \\[0.15em]
$R_p/R_*$ & $0.014751 \pm 0.000059$ & 4 \\[0.15em]
$a/R_*$ & $12.496 \pm 0.015$ & 4 \\[0.15em]
Inc (deg) & $89.183 \pm 0.044$ & 4 \\[0.15em]
Impact parameter & $0.1765 \pm 0.0095$ & 4 \\[0.15em]
Orbital eccentricity $e$ & 0 (fixed) & 5\\[0.15em]
RV semi-amplitude $K$ (m s$^{-1}$) & $1.63 \pm 0.27$  & 5\\[0.15em]
\hline \\[-0.7em]
\multicolumn{3}{l}{\emph{Planetary Parameters}} \\
$R_p (\rearth)$ & $1.478 \pm 0.019$ & 4 \\[0.15em]
$M_p (\mearth)$ & $4.02 \pm 0.68$  & 5 \\[0.15em]
$\rho_p$(g cm$^{-3}$) & $6.88 \pm 1.18$ & 5 \\[0.15em]
$\log g_p$ (cgs) & $3.26 \pm 0.07$ & 5 \\[0.15em]
$a$ (AU) & $0.053 \pm 0.002$ & 5 \\[0.15em]
$T_{\rm eq}$ (K)\tablenotemark{b} & $1037 \pm 13$ & 4 \\[0.15em]
\hline \\[-0.7em]
\textbf{Kepler-93c (companion)} & \\
\multicolumn{3}{l}{\emph{Fit Parameters}} \\
Acceleration (m s$^{-1}$ yr$^{-1}$) & $12.0 \pm 0.4$ & 5\\[0.15em]
\hline \\[-0.7em]
\multicolumn{3}{l}{\emph{Companion Limits}}\\
Mass ($M_J$) & $> 8.5$ & 5 \\
Orbital period $P$ (yr) & $ > 10$ & 5
\enddata
\label{tab:params}
\tablenotetext{}{\textbf{References:}  (1)~\citet{hog_et_al1998}, (2)~\citet{hog_et_al2000}, (3)~\citet{brown_et_al2011}, (4)~\citet{ballard_et_al2014}, (5)~This Paper.} 
\tablenotetext{a}{Systemic velocity at BJD 2456461.57573945.}
\tablenotetext{b}{Assuming a Bond albeo of 0.3.}
\end{deluxetable}                                                                                                                                                                                                  

\begin{figure}[tbp] 
\begin{center}
\centering
\includegraphics[width=0.5\textwidth]{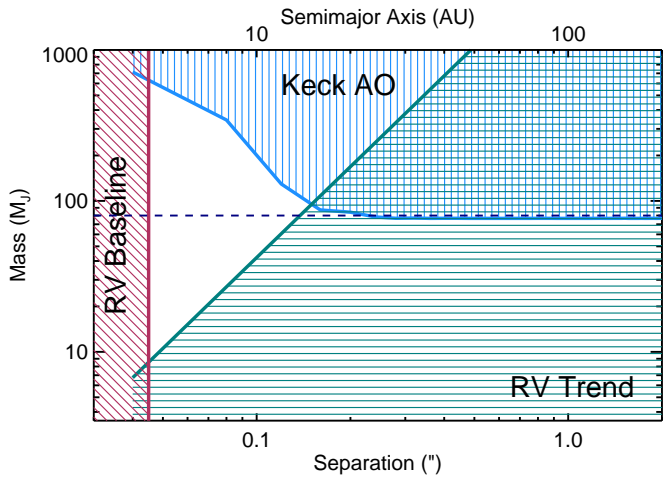}
\end{center}
\caption{Limits on the mass and separation of the companion Kepler-93c. The Keck AO observations exclude a companion within the blue region. The combined HIRES and HARPS-N RVs exclude the teal region due to the amplitude of the trend and the maroon region due to the baseline of the observations. Kepler-93c is therefore constrained to lie within the white region. The dashed purple line divides substellar and stellar companions. These limits assume that the companion has an orbit with $i = 90^\circ$ and $e = 0$. }
\label{fig:93c}
\end{figure}

\begin{figure*}[tbp] 
\begin{center}
\centering
\includegraphics[width=1\textwidth]{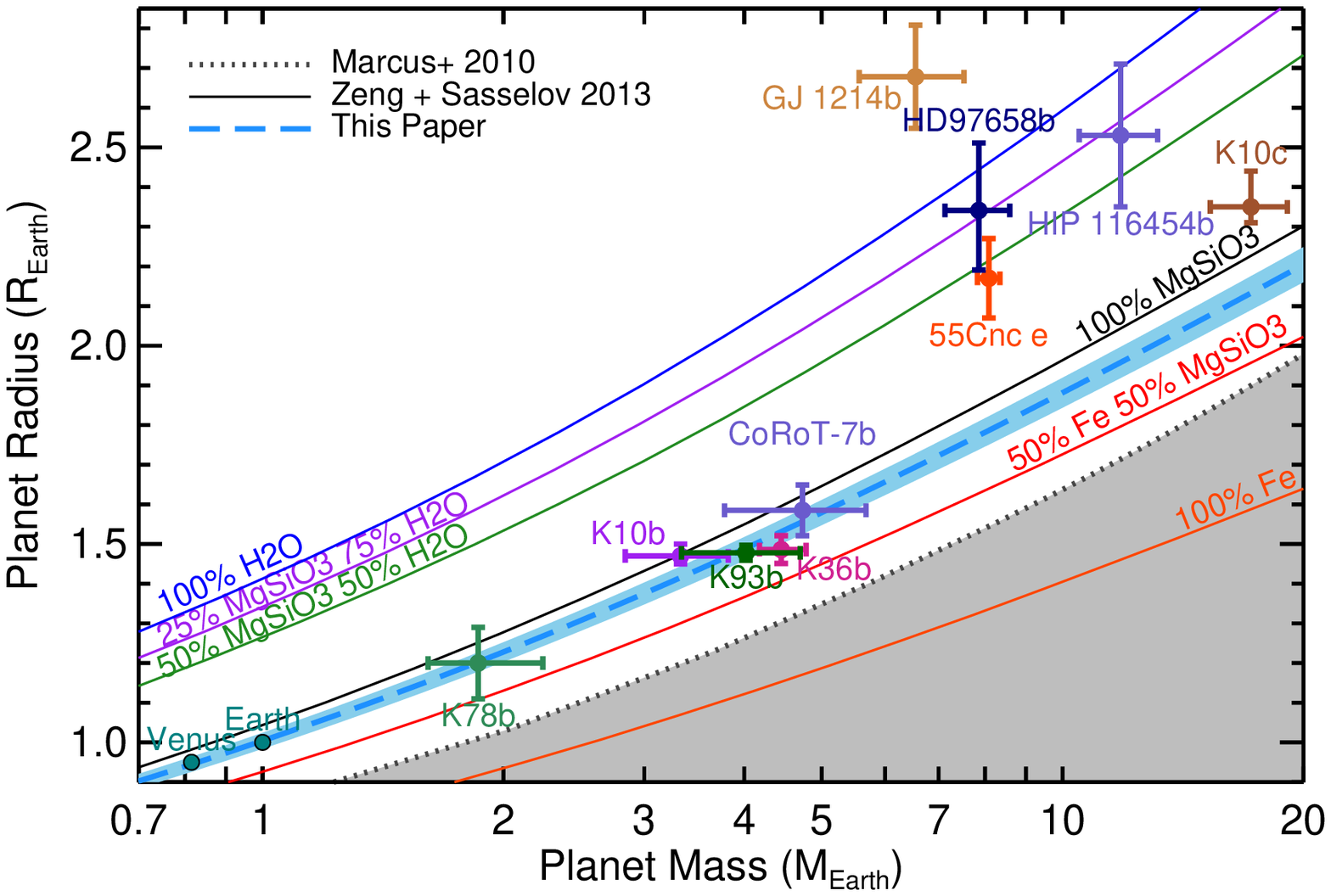}
\end{center}
\caption{Mass-radius diagram for planets smaller than $2.7\rearth$ with masses measured to better than 20\% precision. The shaded gray region in the lower right indicates planets with iron content exceeding the maximum value predicted from models of collisional stripping \citep{marcus_et_al2010}. The solid lines are theoretical mass-radius curves \citep{zeng+sasselov2013} for planets with compositions of 100\% H$_2$O (blue), 25\% MgSiO$_3$ -- 75\% H$_2$O (purple), 50\% MgSiO$_3$ -- 50\% H$_2$O (green), 100\% MgSiO$_3$ (black), 50\% Fe -- 50\% MgSiO$_3$ (red), and 100\% Fe (orange). Our best-fit relation based on the \citet{zeng+sasselov2013} models is the dashed light blue line representing an Earth-like composition (modeled as 17\% iron and 83\% magnesium silicate using a fully-differentiated, two-component model). The shaded region surrounding the line indicates the 2\% dispersion in radius expected from variation in Mg/Si and Fe/Si ratios \citep{grasset_et_al2009}.} 
\label{fig:comp}
\end{figure*}

The expected circularization timescale for Kepler-93b is significantly shorter than the $6.6\pm0.9$~Gyr age of the star \citep{ballard_et_al2014}. Following \citet{goldreich+soter1966}, we calculated a  tidal circularization timescale of \tcircest for a $4.02~\mearth$, $1.48~\rearth$ planet in an orbit with $a = 0.053~AU$ around a $0.91\msun$~star. We assumed $Q = 100$ based on the tidal quality factors estimated for terrestrial planets in the Solar System \citep{yoder1995, henning_et_al2009}. Obtaining a tidal circularization timescale similar to the age of the system would require $Q = 9000$, comparable to the estimate for Neptune \citep{zhang+hamilton2008}. Although the tidal circularization argument is consistent with the preference for a circular orbit, we caution that the tidal quality factors for exoplanets are largely unknown. 

\subsection{Limits on the Properties of Kepler-93c}
The baseline of our RV data is too short to measure the period and minimum mass of the perturber responsible for the long-term trend, but we can place lower limits on the companion properties.  \citet{wang_et_al2014c} conducted a similar analysis of the properties of \mbox{Kepler-93c} based on AO observations and Keck/HIRES RVs. They found a linear RV trend of $12.2 \pm 0.2$~m~s$^{-1}$~yr$^{-1}$ and argued that Kepler-93c is most likely to have a mass below $101~M_J$ and a semi major axis $a = 15.5 - 33$~AU if it is a stellar companion. For the substellar case, they found limits of $a = 5.5 - 27.6$~AU and $M = 10 - 80 M_J$. 

Our additional two years of HARPS-N observations have allowed us to further restrict the allowed parameter space for Kepler-93c.  We measured a linear trend of $12.0 \pm 0.4$ m~s$^{-1}$~yr$^{-1}$ for 5~years, implying that \mbox{Kepler-93c} has $P > 10$~yr and $M > 8.5~M_J$. Assuming the 100~pc distance to Kepler-93 estimated by \citet{ballard_et_al2014}, the resulting semimajor axis $a > 4.5$~AU corresponds to an angular separation of $0\farcs045$.  At this separation, the detection limit from Keck AO imaging is 1.7~\emph{Ks} magnitudes fainter than Kepler-93. We can therefore place an upper limit of $Ks >10.1$ on Kepler-93c unless Kepler-93c happened to have an orbital geometry precluding detection at the epoch of the Keck observations. Converting the $Ks$ upper limit into a mass limit via the \citet{delfosse_et_al2000} relation\footnote{The Delfosse relation predicts stellar mass from $Ks_{\rm CIT}$ whereas the Keck observations were acquired in $Ks_{\rm 2MASS}$. We converted between the two systems assuming a color of $J-K = 1$ and using the color-dependent conversions provided at \url{http://www.astro.caltech.edu/~jmc/2mass/v3/transformations/}.} and the distance, we found a mass upper limit of $0.64 \msun$ for angular separations beyond $0\farcs045$. We display the combined limits from the AO and RV data in Figure~\ref{fig:93c}. In the future, astrometric measurements from Gaia \citep{perryman_et_al2001} will likely provide additional constraints on the properties of the Kepler-93 system. We will then be able to investigate the dynamical history of the system and test whether Kepler-93c might be responsible for scattering Kepler-93b inward onto a short-period orbit. 

\section{Discussion and Conclusions}
\label{sec:disc}
Combining our estimate of \massest for the mass of Kepler-93 with the radius estimate of $Rp = 1.478 \pm 0.019 \rearth$ from \citet{ballard_et_al2014}, we find a density of \rhoest. In Figure~\ref{fig:comp}, we show \mbox{Kepler-93b} on the mass-radius diagram. In this diagram we plot only those planets smaller than $2.7\rearth$ and with masses determined to a precision better than 20\%. In addition to Venus and the Earth, there are ten such planets. We observe that Kepler-93b falls in a cluster of planets with radii 50\% larger than that of the Earth, all of which have extremely similar densities: \mbox{Kepler-10b} \citep[$\rho = 5.8 \pm 0.8$~g~cm$^{-3}$;][]{dumusque_et_al2014}, \mbox{Kepler-36b} \citep[$\rho = 7.46^{+0.74}_{-0.59}$~g~cm$^{-3}$;][]{carter_et_al2012}, and CoRoT-7b \citep[$\rho = 6.56 \pm 1.40$~g~cm$^{-3}$;][]{barros_et_al2014, haywood_et_al2014}. This cluster falls upon a relation that includes Earth, Venus, and Kepler-78b \citep{howard_et_al2013, pepe_et_al2013}, which is itself only 20\% larger than the Earth. To investigate this further, we used the two-component iron-magnesium silicate models of \citet{zeng+sasselov2013} to see if we could find a single composition that explained these seven worlds.  

For the solar system planets, we artificially include mass and radius errors equal to the mean fractional errors for the exoplanets considered so that they do not have undue influence on the resulting fit. We find the lowest $\chi^2$ for a model composition of 83\%~MgSiO$_3$ and 17\%~Fe. We arrive at the same best-fit relation when we exclude Earth and Venus. We caution that the two-component models used in this analysis make two simplifying approximations about the interior structure of planets that cause the core mass fraction to be underestimated: (1) the core contains only iron and the mantle contains only magnesium silicate and (2) the planet is completely dry with no water content. Accordingly, we expect the actual core mass fraction to be slightly higher by 5-8\% to account for incorporation of lighter elements like oxygen, sulfur, and silicon in the core and the inclusion of water in the mantle. In addition, there could be a change of roughly 2\% towards higher or lower core fractions due to uncertainties in the equations of state used in the model calculations. Our purpose in this exercise is to test whether we can find one composition that successfully explains all seven planets, not to place stringent constraints on the abundance of magnesium silicate or iron. 

Intriguingly, all of these planets, which are smaller than $1.6\rearth$, have a tight dispersion around this best-fit compositional curve, suggesting that the distribution of small planet compositions has low intrinsic scatter. In the solar system, the strong agreement between abundance ratios of elements in meteorites and those of the solar photosphere \citep{lodders2003} is a key constraint by which we deduce the composition of the interior of the Earth. Therefore, we might look to the bulk abundances of exoplanet host stars for similar constraints on the interior compositions of their terrestrial planets. \citet{grasset_et_al2009} use a set of planetary models to investigate the dependence of planet radii on elemental abundances. Varying the ratios of iron to silicate and magnesium to silicate within the range observed for the photospheric abundances of nearby exoplanet host stars \citep{beirao_et_al2005, gilli_et_al2006}, \citet{grasset_et_al2009} predicted that the radii of terrestrial planets would vary by roughly 2\% at a given mass. Our findings are in agreement with this picture: We measure a mean absolute deviation of 1.9\% between the estimated planet radii and the values predicted by a 83\%~MgSiO$_3$/17\%~Fe model for planets less massive than $6\mearth$. Indeed, rocky planets very close to their host stars seem to obey a well-defined relationship between radius and mass, although with only 5 such examples outside the Solar system, the immediate task is to characterize other terrestrial exoplanets with similar precision. Increasing the sample of small planets with well-constrained masses and radii will allow us to learn whether additional rocky planets could also be explained by a single mass-radius relation and investigate whether the relation found for close-in planets extends to planets in more distant orbits.

Our mass-radius diagram also includes five planets more massive than $6\mearth$:  55 Cnc e, GJ1214b, HD97658b, HIP 116454b, and Kepler-10c. In contrast, none of these more massive planets have a high density consistent with the best-fit magnesium silicate/iron composition described above. In agreement with \citet{rogers2014}, we find that planets larger than approximately $1.6\rearth$ (e.g., more massive than approximately $6\mearth$) contain significant fractions of volatiles or H/He gas. These planets appear to have a diversity of compositions that is not well-explained by a single mass-radius relation \citep{wolfgang+lopez2014}.

The discussion above focused exclusively on planets smaller than $2.7\rearth$ with masses measured to better than 20\%. Some low-mass worlds with very low densities are known, notably the Kepler-11 system \citep{lissauer_et_al2013} and KOI-314c \citep{kipping_et_al2014b}. Thus we are not proposing that all planets less massive than $6\mearth$ obey a single mass-radius relation; rather, we suggest that the rocky analogs of the Earth might do so.

\acknowledgments
The HARPS-N project was funded by the Prodex Program of the Swiss Space Office (SSO), the Harvard- University Origin of Life Initiative (HUOLI), the Scottish Universities Physics Alliance (SUPA), the University of Geneva, the Smithsonian Astrophysical Observatory (SAO), and the Italian National Astrophysical Institute (INAF), University of St. Andrews, Queen's University Belfast and University of Edinburgh. The research leading to these results has received funding from the European Union Seventh Framework Programme (FP7/2007-2013) under Grant Agreement No. 313014 (ETAEARTH).
C. D.  is supported by a National Science Foundation Graduate Research Fellowship. X. D. would like to thank the Swiss National Science Foundation (SNSF) for its support through an Early Postdoc Mobility fellowship. P.~F. acknowledges support by  Funda\c{c}\~ao para a Ci\^encia e a Tecnologia (FCT) through Investigador FCT contracts of reference IF/01037/2013 and POPH/FSE (EC) by FEDER funding through the program ``Programa Operacional de Factores de Competitividade - COMPETE''. This publication was made possible through the support of a grant from the John Templeton Foundation. The opinions expressed in this publication are those of the authors and do not necessarily reflect the views of the John Templeton Foundation.

\bibliography{../mdwarf_biblio.bib}
\clearpage

\end{document}